\documentclass[preprint,secnumarabic,nobalancelastpage,nofootinbibt]{revtex4}

\usepackage{graphics}      % standard graphics specifications
\usepackage{graphicx}      % alternative graphics specifications
\usepackage{longtable}     % helps with long table options
\usepackage{url}           % for on-line citations
\usepackage{bm,color}            % special 'bold-math' package
\usepackage{amssymb, amsmath, enumerate, theorem, epsfig,subfigure,setspace}
\usepackage{amsfonts}

\newcommand{\nb}[1]{\textcolor{black}{#1}}

\begin{document}

\title{ Networks of non-equilibrium condensates for global optimization}
 \author{Kirill P. Kalinin$^1$ and Natalia G. Berloff$^{2,1}$ }
\email[correspondence address: ]{N.G.Berloff@damtp.cam.ac.uk}
\affiliation{$^1$Department of Applied Mathematics and Theoretical Physics, University of Cambridge, Cambridge CB3 0WA, United Kingdom}
\affiliation{$^2$Skolkovo Institute of Science and Technology Novaya St., 100, Skolkovo 143025, Russian Federation}

\date{10 August 2018}

\begin{abstract}{Recently several gain-dissipative platforms based on the networks of optical parametric oscillators, lasers and various non-equilibrium Bose--Einstein condensates have  been proposed and realised  as analogue Hamiltonian simulators for solving large-scale hard optimisation problems. However,  in these realisations the parameters of the  problem depend on the node occupancies that are not known {\it a priori}, which limits the applicability of the gain-dissipative simulators  to the classes of problems easily solvable by  classical computations. We show how to overcome this difficulty  and formulate  the principles of operation of such simulators for solving the NP-hard large-scale optimisation problems such as constant modulus continuous quadratic optimisation and quadratic binary optimisation for any general matrix.  To solve such problems any gain-dissipative  simulator has to implement a feedback mechanism for the dynamical adjustment of the gain and coupling strengths.}%  The estimates of the time operation of the physical implementation of the gain-dissipative simulators for  large matrices show the speed-up of the several orders of magnitude  in comparison with classical computations.  }  
%We propose such models as a stand along algorithms and as a benchmark for the performance testing of the gain-dissipative simulators.} % and discuss the signatures of a quantum speedup in such simulators.}
\end{abstract}

\maketitle

In the last five years we have seen the rapid emergence of a new field at the intersection of laser and condensed matter physics, engineering and complexity theories which aims to  develop quantum devices to simulate classical  spin problems faster than on classical von Neumann architecture using a gain-dissipative principle of operation.    If the simulated spin problem is NP-hard, then solving it efficiently opens a new route for solving  many practically relevant NP problems \cite{toby}. \nb {Certainly, solving such problems by any physical means would  require the number of operations that grows exponentially fast with the size of the problem (the `NP-hardness assumption' \cite{aaronson05}), but finding the ways to solve these problems for a fixed problem size faster than by a classical Turing machine has a considerable practical importance. }  Among such strongly NP-hard problems are \nb {finding the global minimum of the Ising and XY Hamiltonians} \cite{ZHANG06}, for which even an approximate solution is hard to find \cite{APX}. Several  platforms were proposed and demonstrated as a  proof-of-principle for finding the global minimum of such spin Hamiltonians:  injection-locked laser systems \cite{yamamoto11},
the networks of optical parametric oscillators, \cite{yamamoto14, yamamoto16a, yamamoto16b,takeda18},  coupled lasers \cite{coupledlaser}, polariton condensates \cite{NatashaNatMat2017}, and photon condensates \cite{KlaersNatPhotonics2017}. The main principle of such simulators is based on a gain process that raises the system above the threshold for a phase transition to a coherent state  (a Bose-Einstein condensate, laser coherence, synchronisation of oscillators, etc.).
Since the threshold is first reached at the state that maximises the occupation for a given pumping intensity, \nb { this state  is related to a ground state of a particular spin Hamiltonian \cite{ohadi16, NatashaNatMat2017}}. 
%Accordingly, we refer to these platforms as gain-dissipative (GD) analog Hamiltonian simulators  \cite{blockchain} that, in spite of  having  different quantum hardwares,  share the basic principle that suggests the convergence to the global minimum of the Ising or XY Hamiltonian. 

Such Hamiltonians are written for $N$ classical `spins' ${\bf s}_j=(\cos \theta_j, \sin\theta_j).$ In the XY model `spins' are continuous, whereas for the Ising model the `spins' take discrete values as the phases $\theta_j$ are restricted to $0$ or $\pi$.  The `spins' are coupled to each other with the `strengths' $J_{ij}$ that can be positive (ferromagnetic) or negative (antiferromagnetic), so 
that at the ground state   `spins' arrange their orientation as to minimise  $H=-\sum_{i=1}^N \sum_{j=1}^N J_{ij} \cos(\theta_i -\theta_j) $.  Finding the global minimum of the Ising or XY Hamiltonians are also known as the quadratic binary optimisation and constant modulus  quadratic continuous optimisation problems, respectively.

In the gain-dissipative simulators the `spin' -- the node (the `bit')  of the simulators -- is represented by the phase of a condensate  at a particular spatial position \cite{NatashaNatMat2017,  KlaersNatPhotonics2017} or by the phase of a coherent
state generated in a laser cavity \cite{takeda18,coupledlaser}, so by the phase of the so-called {\it coherent center} (CC).
All of the proposals to use the gain-dissipative simulators  to find the absolute minimum of the spin Hamiltonians  suffer from a serious limitation. As we show below, the coupling strengths $J_{ij}$ in such systems  are modified by the occupations (number densities) of CCs  $i$ and $j$. The densities, however, are not known {\it a priori} and, for a general matrix ${\pmb J}$  will be different from one CC to another. The density inhomogeneity may also lead to the system fragmentation when different subsystems acquire their own mode of oscillation, and the coherent steady state across all CCs will never be reached.  In all the previous experimental realisations of the gain-dissipative simulators, an explicit or implicit assumption was made that the coupling terms are small
so that each laser or condensate  is stabilised independently at the same
steady-state amplitude \cite{yamamoto11, yamamoto14,takeda18,coupledlaser,NatashaNatMat2017}. Such  assumption is justified only for the simplest structures of the coupling matrix ${\pmb J},$ where all CCs are almost equally connected  with about the same coupling strengths. The found solution for a more general matrix is bound to be either only  approximate or invalid. The problem of unequal densities has been recognised before \cite{lelei17}, but the proposed method to reduce the
heterogeneity in densities was to drive the system using randomly modulated signals and  can yield  only very modest improvement and only if the densities are quite close for the unmodulated signals. 

In this paper we formulate the technological requirements for gain-dissipative platforms to be used as analogue Hamiltonian optimisers. We develop a general framework for the operation of the gain-dissipative analogue simulators based on the Langevin gain-dissipative equations written for  a set of CCs. We derive the rate equations for the geometrically coupled CCs such as in polariton or photon condensates. We show that by establishing a feedback connection between the gain mechanism and the density of the CC we can drive the system to the coherent ground state of the  XY model, while the minimisers of this ground state will give the true minimum  precisely for the externally provided coupling strengths.  This framework allows us to formulate the hardware requirements for a physical realisation of any such  simulator to achieve such a minimum and argue that  such requirements are within the recent technological capabilities. %We expect that for large problem sizes the analogue simulator would outperform the classical computations by several orders of magnitude. 

\begin{figure}[t!]
\centering
  \includegraphics[width=8.cm]{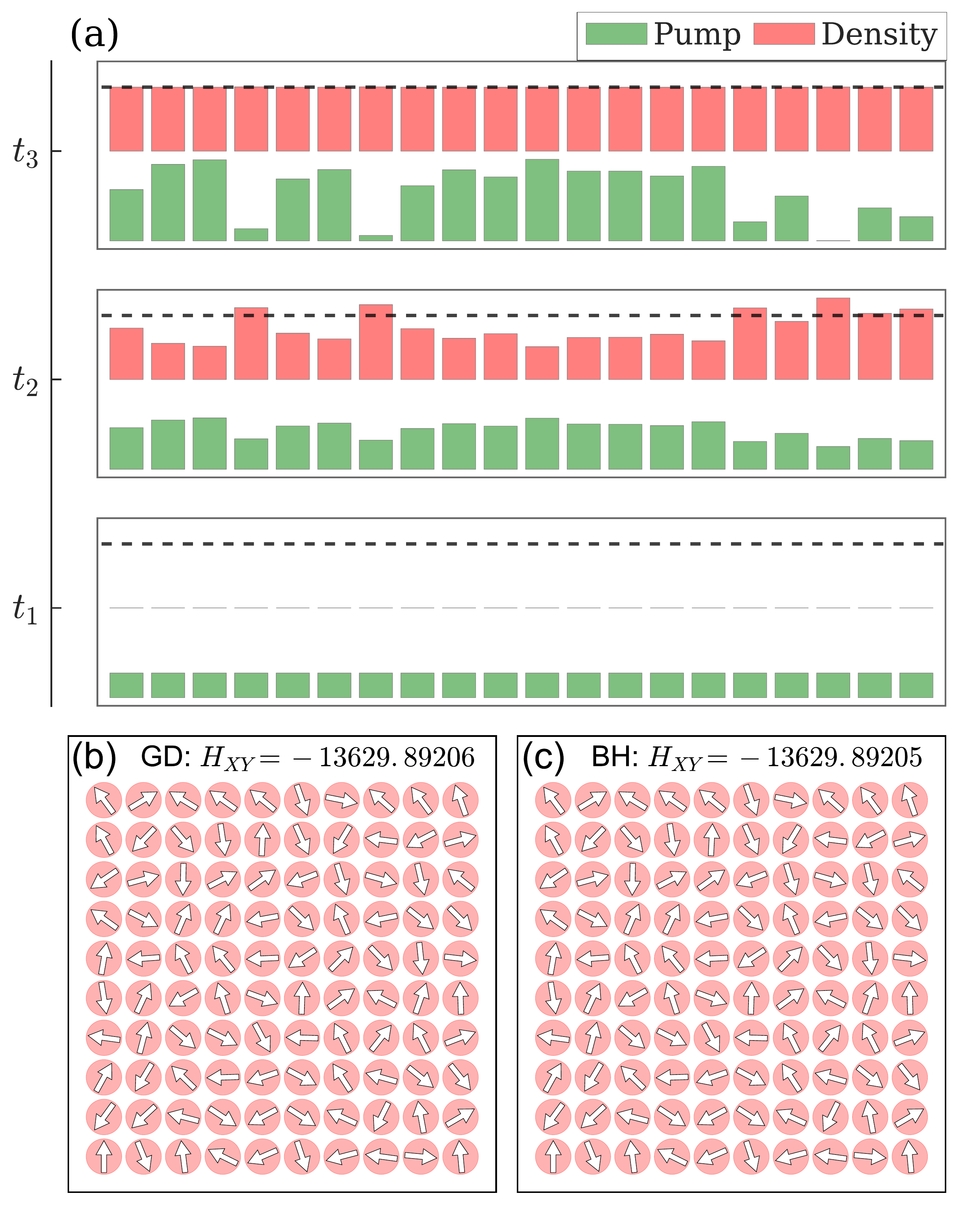}
%\vspace{-3cm}      	
\caption{ (a) The schematics  of operation of the GD simulator. Initially, at $t=t_1$ all CCs are equally pumped (green bars) below the threshold, and all have negligible number densities. As the pumping  intensities increase and depending on the connectivity between CCs,  the different CCs emerge  (red bars)  with different number densities as  shown at some $t=t_2$. The individual control of the pumping intensity  as described by Eq.~(\ref{gamma})  leads to the steady state with all the densities reaching the threshold at $t=t_3$. (b-c) The spin configurations of the absolute minimum of XY Hamiltonian for   $N = 100$ CCs  found by (b) numerical evolution of Eqs. (\ref{ai},\ref{gamma},\ref{gamma2})  as described in the main text and (c) by the basin-hoping global optimization algorithm \cite{BasinHopping1997}. All spins are coupled with all the other spins with the coupling strengths   that are randomly distributed in  $[-10,10]$. The values of the objective functions agree to 10 significant digits between two different methods. }
 \label{Figure1}
\end{figure}

 The operation of a gain-dissipative simulator consists of two stages: bosonic stimulation below the threshold and the coherence of operations at and above the threshold. As one ramps up the power of the gain mechanism (e.g. laser intensity) the gain overcomes the linear losses and is stabilised by  the nonlinear gain saturation. The emergent coherent state maximises the total number of particles (minimises losses)  and, therefore, minimises the Ising or XY Hamiltonian depending on whether the phases of  the CCs  \nb {are discrete or continuous}, respectively. To derive the governing equations for the system one can describe each CC  at a position ${\bf r}= {\bf r}_i$ by a classical complex function  $\Psi_i(t).$   
 Depending on the system, the couplings $K_{ij}$ between CCs can have a different origin: they can be geometrically induced by the particle outflow from other CCs \cite{NatashaNatMat2017} or induced by the mutual injection rate between lasers \cite{takeda18, Strogatz2001} or spatially separated condensates; they can be controlled by external potentials  \cite{KlaersNatPhotonics2017}.  \nb {In what follows we derive  the rate equations for the CCs in the geometrically induced flow in the network of non-equilibrium condensates (such as e.g. polariton or photon condensates) starting from the 
 mean--field description by the complex Ginzburg-Landau equation  (cGLE) \cite{goveq, carusotto}. The CC in this case  is the wavefunction of an individual condensate in the lattice where the spatial degree of freedom has been integrated out.  Such  mathematical elimination of the spatial degree of freedom has been done before \cite{Sigurdsson2017} using the infinite quantum well orthogonal basis. We are interested in the time evolution of the complex amplitudes describing the individual condensates in the lattice; therefore,   we will use  a non-orthogonal basis associated with the wavefunction of an individual condensate.} \nb{The cGLE \cite{goveq} is a driven-dissipative equation of non-equilibrium condensates that has been extensively used to model the steady states, spatial pattern formation  and  dynamics of polariton  condensates  (for review see \cite{reviewCarusotto, revKeelingBerloff}). It can be written  as }   
 \begin{equation}
 {\rm i} \frac{\partial \psi}{\partial t} = - \nabla^2 \psi + ( \tilde U - i  \tilde \sigma) |\psi|^2 \psi +  {\rm i} \biggl(\sum_{i=1}^NP(|{\bf r}-{\bf r}_i|) f_i(t)-\gamma_c\biggr) \psi,
 \label{gpe}
 \end{equation}
where $\psi({\bf r},t)$ is the wavefunction of the system,  $\tilde U$ is the strength of the \nb {delta-function} interaction  potential, the sum on the right-hand side represents the rate of adding particles in $N$ spatial locations ${\bf r}={\bf r}_i$, $i=1,...,N$, $P_i({\bf r})\equiv P(|{\bf r}-{\bf r}_i|)$ is a given spatially localised pumping profile, that creates the condensate with a wavefunction $\phi_i({\bf r})=\phi(|{\bf r}-{\bf r}_i|)$ centred at ${\bf r}={\bf r}_i$ with a normalised number density  so that $2\pi \int |\phi(r)|^2 r\, dr=1$. Furthermore,  $f_i$ is the time-dependent part of the pumping at ${\bf r}={\bf r}_i$,  $\gamma_c$ is the rate of linear losses and $ \tilde \sigma$ is the rate of the density-dependent losses. In writing Eq.~(\ref{gpe}) we let $\hbar=1$ and $m=1/2$.  If the distances between CCs are larger than the width of  $P(r)$, the wavefunction of the condensate can be written as  
$
\psi({\bf r},t) \approx \sum_{i=1}^N a_i(t) \phi_i({\bf r})
$ \nb{\cite{PO}},
where $a_i(t)$ is the time-dependent complex amplitude. 
We substitute the expression for $\psi$ into Eq.~(\ref{gpe}), multiply by $\phi_j^*$ for $j=1,...,N$, and integrate in space to yield the rate equations for $a_j(t)$
\begin{equation}
i \dot{\pmb a}^T=\left[- {\pmb D}+ i {\pmb \Gamma}-i\gamma_c {\pmb \Xi}+( \tilde U - i \tilde \sigma) {\pmb Q}  \right] {\pmb a}^T {\pmb \Xi}^{-1},
\label{matrix}
\end{equation} 
where $ {\pmb a}=\{a_i\},$  $ {\pmb \Xi}=\{\chi_{ij}\},$ $ {\pmb D}=\{d_{ij}\}$, ${\pmb \Gamma}=\{\Gamma_{ij}\}$,  ${\pmb Q}=\{ q_{ij}\},$
and
$\chi_{ij}= \int \phi_i\phi_j^*\, d{\bf r}$,   $q_{ij}= \int |\sum_{k = 1}^N a_k \phi_k|^2 \phi_i \phi_j^* d{\bf r}$,  $\Gamma_{ij}=\sum_k f_k\int P_k \phi_i \phi_j^*\, d{\bf r},$ and 
$d_{ij}\equiv\int \phi_j^*\nabla^2 \phi_i \, d{\bf r}.$
An asymptotics of  $\phi(r)$ for the  Gaussian pumping profile have been developed in \cite{PO} \nb {where we showed that} function $\phi$ can be approximated by 
\begin{equation}
\phi(r)= \sqrt{\frac{2}{\pi}}\beta \exp[-\beta r + i k_c r],
\nb{\label{phi}}
\end{equation}
 where $k_c$ is the outflow velocity and $\beta$ is the inverse characteristic width of the condensate \cite{PO}. 
The integrals $\chi_{ij}=\int  \phi_i \phi_j^*\, d{\bf r}$ for $i\ne j$ can be evaluated in elliptical coordinates in terms of the Bessel functions \cite{matter}
\begin{equation}
\chi_{ij}= 2 \beta^2 l_{ij} \bigg[ \frac{1}{\beta} J_0 (k_c l_{ij}) K_1 (\beta l_{ij}) 
+  \frac{1}{k_c} J_1 (k_c l_{ij}) K_0 (\beta l_{ij})\bigg],
\label{bessel}
\end{equation}
 where $l_{ij}=|{\bf r}_i - {\bf r}_j|$. 
  We assumed that the CCs are well separated: $l_{ij}\beta \gg 1,$ so that for $i\ne j$ we have $\chi_{ij} \ll  1$ as follows from Eq.~(\ref{bessel}). The integrals in the off-diagonal terms in ${\pmb Q}$  are of the same order in smallness as $\chi_{ij}$. \nb {Below the threshold $a_i(t)=0$ and at the   threshold $|a_i(t)|^2$ are non-zero but small (polaritons are bosons and therefore obey the Bose-Einstein statistics at low densities characterised by $a_i(t)$).} The off-diagonal terms in ${\pmb Q}$ are quadratic in these small quantities and can be neglected to the linear order in Eq. (\ref{matrix}), so that ${\pmb Q}$ is the diagonal matrix with elements $|a_i|^2q$, where $q=2 \pi \int |\phi|^4r\, dr$. One can show that  $d_{ij}\ll d_{ii}$  and can be neglected to the linear order, so that ${\pmb D}=d {\pmb I}$, where $d=2 \pi \int \phi\nabla^2 \phi r \, dr$. The matrix ${\pmb \Xi}$ has $1$ on the diagonal and $\chi_{ij}$ as $ij-$th element. Taking the inverse and keeping only up to the linear order terms gives a matrix with 1 on the diagonal and $-\chi_{ij}$ as the $ij-$th element. Finally, to the linear order terms the elements of matrix ${\pmb \Gamma}$   are $\Gamma_{ii}=f_i p$  and $\Gamma_{ij}=(f_i p_{ij} +f_j p_{ji}^*)$ for $i\ne j$, where $p=2 \pi\int P(r) |\phi|^2 r\, dr,$ and $p_{ij}=\int P_i \phi_i \phi_j^*\, d{\bf r}$. To the linear order terms in small quantities $|a_i(t)|^2$ and $p_{ij}$ and neglecting $\chi_{ij} \ll p_{ij}$ for $i\ne j$ Eqs. (\ref{matrix}) become 
\begin{equation}
 \frac{d\Psi_i}{dt}= \Psi_i (\gamma_i^{\rm inj}  - \gamma_c-({\rm i}U+\sigma)|\Psi_i|^2) + \sum_{j, j\ne i}^N \Delta_{ij}^{\rm inj}K_{ij}\Psi_j +D\xi_i(t)
,
\label{ai}
\end{equation}   
where we  used the quadratic smallness of $p_{ij} - p_{ji}^*$, let $\Psi_i=a_i \exp({\rm i}d t)$, $\gamma_i^{\rm inj}=f_i p$, $U=q \tilde U$, $\sigma = q \tilde \sigma$, $K_{ij}= Re\{p_{ij}\}/p$, $\Delta_{ij}^{\rm inj}=\gamma_i^{\rm inj}(t) + \gamma_j^{\rm inj}(t)$ and introduced the Langevin noise $\xi_i(t)$ ($\langle \xi_i(t) \xi^*_i(t')\rangle=\delta(t-t')$) which represents  intrinsic vacuum fluctuations and classical noise with a diffusion coefficient $D$ which disappears at the threshold. 
 In other platforms such as the Coherent Ising Machines \cite{yamamoto14} the injection does not have to be symmetric between the nodes -- this will be modelled by introducing an asymmetry parameter $\delta$ so that $\Delta_{ij}=\gamma_i^{\rm inj}(t) + (1-\delta) \gamma_j^{\rm inj}(t)$, where $\delta=0$ for  symmetrically coupled CCs. Equations (\ref{ai}) are the rate equations on the CCs coupled with the strengths $\Delta_{ij}K_{ij}.$  Note, that by writing the coupling strength in this form we separated the effect of what is  not known a priori (pumping intensity, energy at the threshold etc) from $K_{ij}$ that depends on the characteristics of the system that are known and for geometrically coupled condensate, for instance, depend on the distance between CCs $i$ and $j$. 
 %As one raises the injection rates $\gamma_i^{\rm inj}$ from zero the phases of the CCs, $\theta_i$, align to minimize the XY Hamiltonian.  This assumption has been at the core of all the proposals to use the gain-dissipative simulators as global optimizers. 
 To show how Eqs.~(\ref{ai}) lead to the XY model minimisation we rewrite them in terms of the number densities $\rho_i$ and phases $\theta_i$ using the Madelung transformation $\Psi_i=\sqrt{\rho_i} \exp[{\rm i} \theta_i]$ while suppressing noise:
 \begin{eqnarray}
\frac{1}{2}\dot{\rho}_i(t)&=&(\gamma_i -\sigma \rho_i) \rho_i + \sum_{j;j\ne i} \Delta_{ij}^{\rm inj}K_{ij} {\sqrt{\rho_i\rho_j}}\cos\theta_{ij},\label{rho}\\
\dot{\theta}_i(t)&=&-U\rho_i -\sum_{j;j\ne i} \Delta_{ij}^{\rm inj}K_{ij} {\frac{\sqrt{\rho_j}}{\sqrt{\rho_i}}} \sin\theta_{ij},\label{theta}
\end{eqnarray}
where $\theta_{ij}= \theta_i-\theta_j$.
  The first term on the right-hand side of Eq. (\ref{theta}) tends to provide $\theta_i$ with its own frequency of oscillations \nb {whereas the second term couples the phases to each other and so   tends to synchronise them, in the full analogy with the Kuramoto model that Eq. ~\ref{theta} represents \cite{kuramoto}. The phase synchronisation  in this context assumes the constant (but not necessarily zero) phase differences between the CCs. }  Phase synchronisation in such a system has been extensively studied  especially in the context of semiconductor laser arrays  \cite{CoupledLasers95, Mandel2000}.  To guarantee that the steady state of the system is reached and coincides with  a minimum of the XY model, however,  one needs to ensure that  the gain mechanism is chosen so that all densities $\rho_i$ are the same \nb{at the steady state at the threshold}. Only under this condition does \nb{the second term on the right-hand side of  Eq. (\ref{theta})} describe the gradient decent to the minimum of the XY model. Previously, any realisation of the XY model using the laser systems (or non-equilibrium condensates) was based on the assumption that all lasers (condensates) have the same steady-state photon (particle) number \cite{CoupledLasers95, Mandel2000,takeda18}. This limits the problems that can be addressed by such a framework to trivial ones where all CCs have an almost equal number of connections   with almost the same pumping rate. \nb{Different densities at the threshold even if the steady state and therefore phase synchronisation is achieved imply that although the system reached the minimum of  the XY Hamiltonian the coupling coefficients $ \Delta_{ij}^{\rm inj}K_{ij} $ are replaced by $ \Delta_{ij}^{\rm inj}K_{ij}  \sqrt{\rho_j/\rho_i}$ where $\rho_i$ and $\rho_j$ are not known {\it a priori}.} To implement any  couplings and connectivities, one needs to be able to control the pumping rate of the individual CCs and bring all the densities to the same value at the threshold, \nb {so that the term $\sqrt{\rho_j/\rho_i}$ is cancelled out.}  We schematically illustrate the operational principle of such control mechanism in Fig.~\ref{Figure1}a. Starting from below the threshold, all CCs are equally pumped  as shown at some $t=t_1$. Depending on the node connectivity, the non-zero densities emerge at different rates for each CC as the pumping intensity increases and takes some CCs above the specified threshold $\rho=\rho_{\rm th}$  as illustrated on Fig.~\ref{Figure1}a at some later time $t=t_2$. The pumping mechanism must be adjusted for each  CC to enable the saturation at the same density: decreased for CCs with densities above the threshold and increased for the CCs below the threshold. The feedback mechanism can be implemented via optical delay lines (in a \nb{network of optical parametric oscillators} system), by adjusting the injection via the spatial light modulator (SLM) (in the polariton and photon condensates) or  by electrical injection (e.g. in the polariton lattices \cite{electrical18}). The mathematical description of such a feedback mechanism  is
  \begin{equation}
 \frac{d \gamma_i^{\rm inj}}{dt} = \epsilon (\rho_{\rm th} - \rho_i),
 \label{gamma}
 \end{equation}
 where $\epsilon$ is a parameter that can be tuned to control the speed of approaching the threshold. 
 The fixed point of Eqs. (\ref{rho}-\ref{gamma}) is
$
 \rho_i=\rho_{\rm th}=(\gamma_i^{\rm inj} -\gamma_c+  \sum_{j;j\ne i}  \Delta_{ij}^{\rm inj}K_{ij} \cos\theta_{ij})/\sigma,
$
with the total particle number given by $M=(\sum_i\gamma_i^{\rm inj}-N\gamma_c +  \sum_{i,j;j\ne i}  \Delta_{ij}^{\rm inj}K_{ij} \cos\theta_{ij})/\sigma.$ \nb {Since the pumping of each of the CCs is controlled independently from the others by gradually increasing the gain from below until the threshold density value is achieved, the condensation will first take place at the minimum of $\sum_i\gamma_i^{\rm inj}$ for a given set of coupling coefficients.  Since $M=N \rho_{\rm th}$ and $N\gamma_c$ are fixed, the minimum of  $\sum_i\gamma_i^{\rm inj}$ is achieved at the maximum of $\sum_{i,j;j\ne i}  \Delta_{ij}^{\rm inj}K_{ij} \cos\theta_{ij}$ so at the  minimum of 
 the XY Hamiltonian $H_{XY}=-\sum_{i,j;j\ne i}  \Delta_{ij}^{\rm inj}K_{ij} \cos\theta_{ij}$.}  The minimum is approached from below by gradually raising the pumping to the threshold which facilitates  the achievement of the global minimum as Fig.~\ref{Figure1}a at $t=t_3$ illustrates. 
 
 In the context of the cGLE  appropriate for the description of non-equilibrium condensates the nonlinear dissipation $\sigma$ can be proportional to the pumping intensity $\sigma= c \gamma_i^{\rm inj}$ \nb {\cite{reviewCarusotto}}, where $c$ is a system dependent parameter that for polariton condensates for instance depends on the decay rate of the particles in the hot exciton reservoir. The steady state of Eqs. (\ref{rho}-\ref{gamma}) becomes 
$
 \rho_i=\rho_{\rm th}=(1-\gamma_c/\gamma_i^{\rm inj}+  \sum_{j;j\ne i}  \Delta_{ij}^{\rm inj}K_{ij} \cos\theta_{ij})/c,
$ with $\Delta_{ij}^{\rm inj}=1+\gamma_j^{\rm inj}/\gamma_i^{\rm inj}$. In this case,  at the threshold \nb {each of $1/\gamma_i^{\rm inj}$ (and so $\sum_i 1/\gamma_i^{\rm inj}$)} is maximised, so again the system reaches the  minimum of 
 the XY Hamiltonian.

If one removes the density heterogeneity the global minimum of the XY model will be achieved, but  the coupling terms $\Delta_{ij}^{\rm inj}(t)K_{ij}$ now depend on the  particle injection rates $\gamma_i^{\rm inj}$ that would not be known {\it a priori} if one requested the equal densities at the threshold. Therefore, not only  $\gamma_i^{\rm inj}(t)$ has to be adjusted in time to equalise the densities  using Eq. (\ref{gamma})  but also  the coupling parameters $K_{ij}$ have to be modified  in time to bring the required couplings $J_{ij}$   at the steady state by
\begin{equation}
\frac{d K_{ij}}{d t} = \hat{\epsilon}( J_{ij}-\Delta_{ij}^{\rm inj}K_{ij} ),\label{gamma2}
\end{equation}
where $\hat{\epsilon}$ controls the rate of the coupling strengths adjustment. Since $\hat{\epsilon}\ll \epsilon$ such adjustments do not significantly slow down the  operation of the simulator as they have to be performed much more rarely then adjustments of the gain. Equation (\ref{gamma2}) indicates that the couplings need to be reconfigured depending on the injection rate: if the coupling strength scaled by the gain at time $t$ is lower (higher) then the objective coupling $J_{ij}$, it has to be increased (decreased).  We have verified that  Eqs. (\ref{ai}, \ref{gamma}, \ref{gamma2})   find not just the value of the global minimum of the XY Hamiltonian for a variety of couplings and sizes of the system, but also the minimizers  as Figs. \ref{Figure1}b,c show. 

%To stand such a challenge  any simulator has to meet several   requirements for its operation such as the realisation of independent control of the couplings between  two nodes, coupling beyond the next neighbour, and the feedback mechanism that allows one to adjust the pumping depending on whether or not the corresponding CC is above or below the threshold.

To illustrate the implementation of the density and coupling adjustments and, therefore, realisation of the global minimum of the XY model by the actual physical system, we apply the developed algorithm to the square lattice of polariton condensates experimentally achieved in our previous work \cite{NatashaNatMat2017}. \nb {The required implementation of  feedbacks  for the density (Eq.~\ref{gamma}) and coupling (Eq.~\ref{gamma2}) adjustments  can be experimentally achieved by adjusting the injection rates and reconfiguring the lattice geometry using the SLM. Liquid crystal based SLMs are widely used to generate arbitrarily structured beams of various intensity in optical micro-manipulation devices. We have already demonstrated the ease of  using such SLMs for changing the coupling intensities in polariton graphs \cite{NatashaNatMat2017}. An alternative beam shaping technology  is based on the digital micro-mirror devices that  are capable of much faster  switching rates for  spatially controllable intensity (in excess of 20kHz whereas liquid crystal based SLMs are typically limited by a switching rate on the order of $\sim 100$ Hz due to the viscosity of the liquid crystal \cite{dmd}). Therefore, the proposed feedbacks are well within the technological capabilities of our platform.} 
 Figure \ref{Figure2}a shows the density profile of $45$ polariton condensates that interact by the outflow of the particles from neighbouring CCs. All CCs are equally pumped (with, say, $\gamma^{\rm inj}\equiv \gamma_i^{\rm inj}$) and, therefore, the  CCs away from the margins have the largest occupation -- they are fed by the particles coming from the eight neighbours. The CCs at the margins have the lowest occupation as they interact with only four or five neighbours. Such density heterogeneity between the lattice sites is clearly seen on Fig.~\ref{Figure2}a with the  condensates on the margins being barely visible. The resulting configuration  realises the global minimum of the XY model, but with the coupling strengths between $i-$th and $j-$th condensates given by $2 \gamma^{\rm inj} K_{ij}\sqrt{\rho_i\rho_j}$ with number densities $\rho_i$ and $\rho_j$ that are not known before the system reaches the configuration shown in Fig. {\ref{Figure2}a.  The numerical simulation of the $7\times7$ polariton lattice  shows not only the density variation between the sites in agreement with the experimental result but indicates the  formation of a spin wave state (Fig.~\ref{Figure2}b) which manifests the presence of various couplings in the lattice. To realise the global minimum of the XY model for the given couplings ($J_{ij}$) we need  to implement the feedback mechanisms  described above and that  we illustrate step by step. First, we remove the density heterogeneity by adjusting the gain mechanism described by Eq.~(\ref{gamma}). The resulting pumping profile  is  shown in Fig. \ref{Figure2}c with the corresponding steady state number  densities and phases  in Fig. \ref{Figure2}d.  The spin wave in the presence of equal densities between the lattice sites is due to  the different  pumping intensities, and therefore, different couplings $\Delta_{ij}K_{ij}$ between the CCs across the lattice.  We adjust  $K_{ij}$ according to Eq.~(\ref{gamma2}) by changing the distances between the sites as  Fig. \ref{Figure2}e illustrates. The final steady state has equal densities and equal antiferromagnetic coupling strength between the nearest neighbours with phases alternating between $0$ and $\pi$ to give the expected global minimum of the XY model.  
\begin{figure}[b!]
\centering
  \includegraphics[width=8.cm]{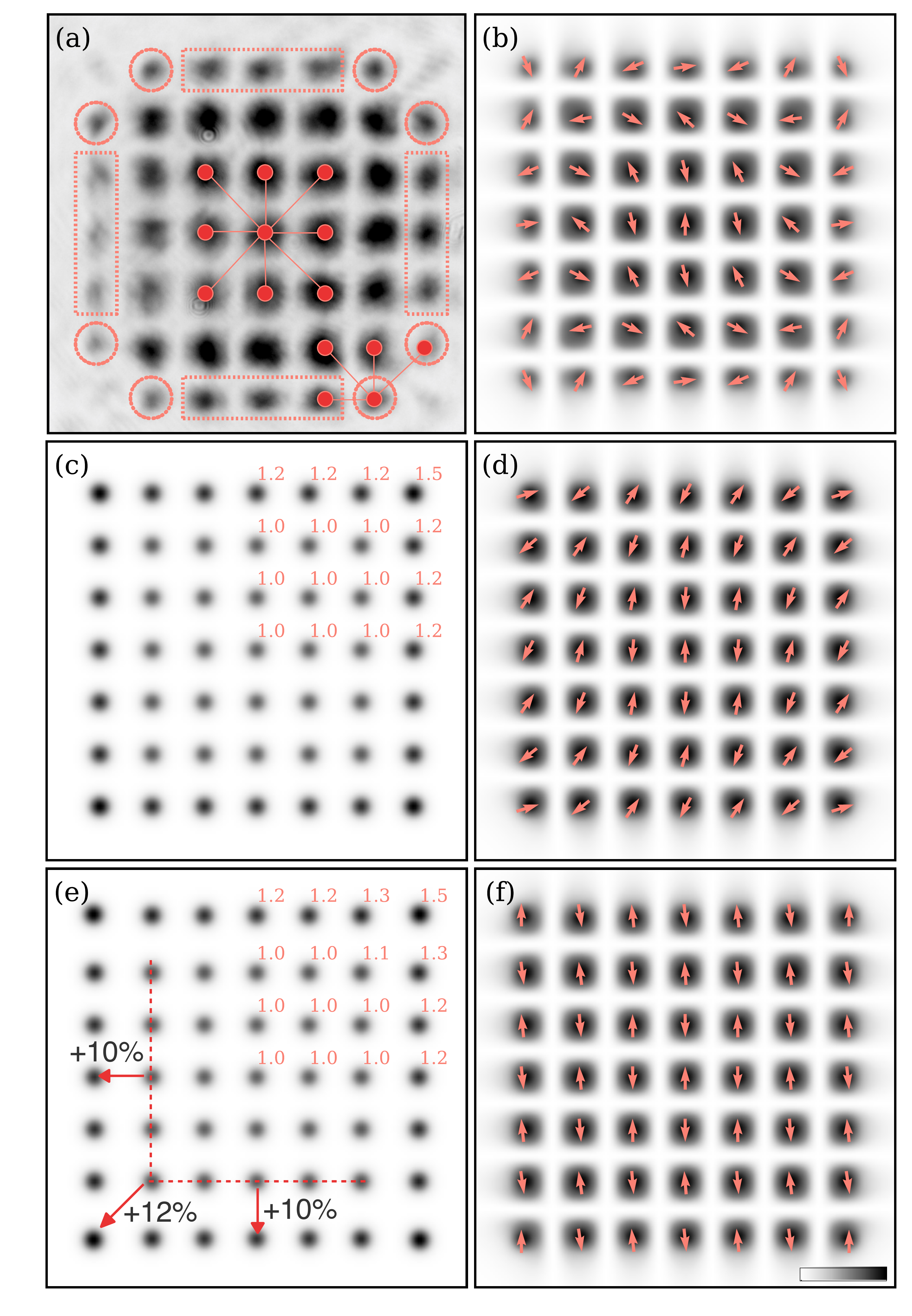}
%\vspace{-3cm}      	
\caption{ (a) Contour plot of the number density of the polariton condensates  formed by non-resonantly pumping polaritons   with equal intensities (reproduced with permission from \cite{NatashaNatMat2017}). The distance between the  neighbours is such that the coupling strength is antiferromagnetic for a polariton dyad with such separation  at the condensation threshold.  Red lines show the particle fluxes between the sites: the central site experiences the inflow of the particles from eight neighbouring sites whereas the  sites on margins have only four or five  neighbours. Dashed figures embrace the condensates with densities lower than the central condensates. (b,d,f) Contour plots of the  steady state number density function $\rho=\int |\psi|^2\, d{\bf r}$  obtained  by the numerical integration of the full dynamical governing equations  for $7\times 7$ lattice and for the parameters used previously \cite{NatashaNatMat2017}. The lattice constant is   $7.5 \mu m$. (c,e) Contour plots of the pumping profiles at the steady state.  Panels (c,d) were obtained by applying the density adjustments according to Eq.~(\ref{gamma}). Panels (e,f) were obtained by applying both the density adjustments and coupling adjustments according to Eq.~(\ref{gamma}) and Eq.~(\ref{gamma2}). The resulting pumping intensities at the lattice sites are  indicated for the top right quarter of (c,e) as the factor  of the pumping at the lattice centre. The coupling strength adjustments are achieved by shifting the lattice sites as shown in red for the bottom left corner only in (e). Figure \ref{Figure2}(a) reprinted with permission from \cite{NatashaNatMat2017}. Published in Nature Materials, Copyright (2017). https://doi.org/10.1038/nmat4971. }
 \label{Figure2}
\end{figure}
%

%So far our discussion has been most directly relevant to the geometrically coupled lattices of lasers or non-equilibrium condensates but the similar ideas of achieving the uniform densities and adjustments of couplings apply to other gain-dissipative simulators such as the Coherent Ising Machine based on the network of degenerate optical parametric oscillators \cite{yamamoto14,yamamoto16a,yamamoto16b}. The Ising Hamiltonian is realised  by incorporating the  continuous spin mapping $ g(x_i)=\frac{1}{2} \alpha x_i^2 - \frac{1}{4} x_i^4$ representing the   double-well potential that imposes spins  discreetness $s_i=x_i/|x_i|, x_i\ne 0$, and $\alpha$ is the system dependent parameter \citep{YamomotoPRE2017}.  To describe such a simulator the equations Eqs.~(\ref{rho}-\ref{theta}) have to be modified as
%\begin{eqnarray}
%\dot{\rho}_i(t)&=&(\gamma_i^{\rm inj}-\gamma_c -\sigma \rho_i) \rho_i + \gamma_i^{\rm inj} \sum \sqrt{\rho_i\rho_j} {w}_{ij} x_i x_j 
%+ \rho_i g(x_i), \label{ising1}\\
%\dot{x}_i(t)&=&-U\rho_i + \gamma_i^{\rm inj} \sum \frac{\sqrt{\rho_j}}{\sqrt{\rho_i}} w_{ij} x_j + g'(x_i).\label{ising2}
%\end{eqnarray}
 % The  system of Eqs. (\ref{ising1}, \ref{ising2}) with the Langevin  noise and together with  the feedback mechanisms given by   Eq.~(\ref{gamma}, \ref{gamma2})  where $K_{ij}=w_{ij}$ and $\Delta_{ij}=\gamma_i^{\rm inj}$ realises the global minimum of the Ising model at the steady state.

%
%
%
The developed procedure for the dynamical adjustment of the gain and coupling strengths, which is reflected in 
Eqs. (\ref{ai},\ref{gamma},\ref{gamma2}) %or Eqs. (\ref{ising1},\ref{ising2},\ref{gamma},\ref{gamma2}), 
could be simulated by a classical computer and  lead to a new class of global optimisation algorithms as we explore elsewhere \cite{kirillAlgorithm} in particular, analysing if or when the actual physical gain-dissipative simulator can  outperform the classical computer algorithm due to inherent parallelism of spanning various phases before the configuration that enables coherence at the threshold  is reached and whether quantum superpositions  contribute to processing of the phase configurations. These results can also be extended to other spin Hamiltonians such as the Ising and Potts \cite{kirillPotts}.

In this paper we discussed gain-dissipative systems such as OPOs, optical cavities, lasers, and non-equilibrium condensates. The physics of the gain-dissipative oscillators may cover a  much large variety of systems such as electronic circuits, voltage controlled oscillators, microelectromechanical systems, spin-torque nano-oscillators, oxide-based oscillators, etc.
Our analysis is applicable to a broad family of oscillatory networks regardless of the nature of each oscillator as long as it is governed by the generic Eqs. (\ref{ai}), can be operated stably at the threshold and allows for the feedback mechanism of adjusting the pumping intensity of an individual CCs and their coupling strengths, similar to the operation of the considered systems.

%Finally, we would like to comment on classical vs quantum operation of such simulators. When a condensate (a coherent state) is formed -- the system behaves classically as many bosons are in the same single-particle mode and non-commutativity of the field operators can be neglected.  However, the condensation process by which the global minimum of the XY model is found involves quantum effects. It was shown before, that the condensation process can be described by a fully classical evolution of the Nonlinear Schr\"odinger equation  that takes into account only stimulated scattering effects and neglects spontaneous scattering \cite{berloffSvistunov}. The classical or quantum assignment to gain-dissipative simulators depends on whether  quantum fluctuations and spontaneous scattering effects during the condensation provide a speed-up in comparison with fully classical noise and stimulated scattering.  This is an important question to address in the future research on such simulators and the comparison with the classical algorithm that we developed based on the gain-dissipative simulators architecture allows one to see if the time to find the solution scales better than with the best classical algorithms. 

%Finally, we would like to conclude with  the comment made by Steiger et al \cite{QuanAnnealPerfomanceTroyer2015}: assuming that there is a quantum speed-up in gain-dissipative simulators we can ask: "Can we mimic the features that produce quantum speedup in an efficient classical code and thus beat the machine again?"
%\section*{Acknowledgements}
NGB acknowledges financial support from the NGP MIT-Skoltech. K.P.K. acknowledges the financial support from Cambridge Trust and EPSRC.

\end{document}